\documentclass[twocolumn,pra,tighten,floatfix,showpacs, amssymb]{revtex4-1}
\usepackage{graphicx}

\def\<{\langle}
\def\>{\rangle}

\def\be{\begin{equation}}
\def\ee{\end{equation}}
\def\Sent{S_{\rm ent}}

\begin{document}
\preprint{cond-mat} \title{Scaling of Entanglement Entropy in Point Contact Free Fermion Systems}

\author{B. Caravan, B. A. Friedman$^\dagger$ and G. C. Levine}

\address{Department of Physics and Astronomy, Hofstra University,
Hempstead, NY 11549}
\address{$^\dagger$Department of Physics, Sam Houston State University, Huntsville TX 77341}

\date{\today}

\begin{abstract}
The scaling of entanglement entropy is computationally studied in several $1\le d \le 2$ dimensional free fermion systems that are connected by one or more point contacts (PC). For both the $k$-leg Bethe lattice $(d =1)$ and $d=2$ rectangular lattices with a subsystem of $L^d$ sites, the entanglement entropy associated with a {\sl single} PC is found to be generically $S \sim L$. We argue that the $O(L)$ entropy is an expression of the subdominant $O(L)$ entropy of the bulk entropy-area law. For $d=2$ (square) lattices connected by $m$ PCs, the area law is found to  be $S \sim aL^{d-1} + b m \log{L}$ and is thus consistent with the anomalous area law for free fermions ($S \sim L \log{L}$) as $m \rightarrow L$. For the Bethe lattice, the relevance of this result to Density Matrix Renormalization Group (DMRG) schemes for interacting fermions is discussed.


\end{abstract}

\maketitle

\section{Introduction}

In a quantum field theory, entanglement entropy of a region $A$ is not a conventional extensive quantity but, rather,  depends upon the bounding surface of the distinguished region $A$, a result known as the entropy-area law \cite{Srednicki,Holzhey,Vidal,Cardy_rev}.  A great deal of work has concentrated on entanglement entropies computed in a homogeneous system; specifically, the bipartite entropy of a $d$-dimensional distinguished region ($A$) separated from the $d$-dimensional bulk by a $(d-1)$-dimensional boundary.  For gapless fermions it has been proven \cite{ferm1,ferm2,ferm3,ferm4,Eisert} that the area law is anomalous and, in contrast to entropy proportional simply to the bounding surface area,
\begin{equation}
\label{area_law}
S = \left(\frac{L}{\epsilon}\right)^{d-1}\ln{\frac{L}{\epsilon}} + O(L^{d-1})
\end{equation}
where $L$ is the linear dimension of the distinguished region, $A$, and $\epsilon$ is a spatial cut-off.


\begin{figure}[ht]
\includegraphics[width=6.0cm]{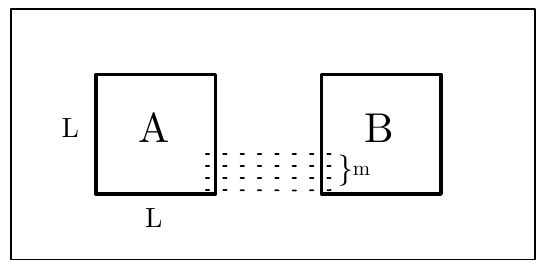}
\caption{\label{fig2} Two 2-d fermion lattices connected by $m$ point contacts.}
\label{lattice}
\end{figure}

However, entanglement entropy should be nonzero for {\sl any} boundary with a dimension, $k$, ranging from $0$ to $d-1$.  In this manuscript, we examine the resulting "sub" area law by studying several noninteracting fermion systems where the codimension of the boundary is variable---albeit in a limited way.  It is known that in the simplest interesting subcase---a zero dimensional boundary (or, point contact) connecting two $1$-dimensional regions---the point contact does does not affect the scaling properties of the entropy of noninteracting fermions other than in the pre factor, and the entropy remains $S \sim \log{L}$ \cite{levine_imp,CenkeXu}. 
\begin{figure}[ht]
\includegraphics[width=5cm]{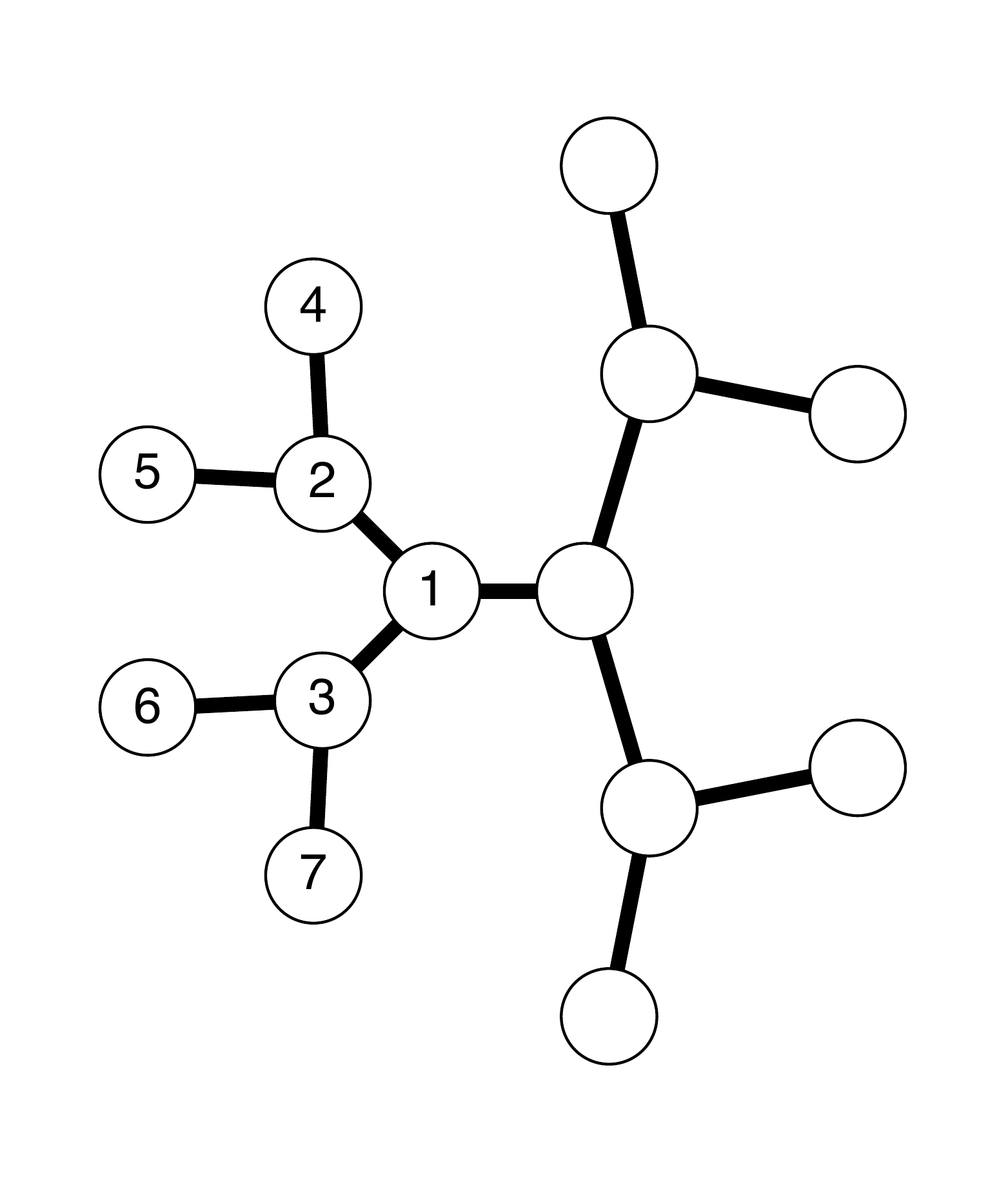}
\caption{A $14$ site $3$ branch bond centered Bethe cluster.}
\label{b1}
\end{figure}

In contrast, for both the Bethe lattice $(d=1)$ and $d=2$ lattices, we find that a single point contact (PC) connecting two $L^d$ site systems generates an entropy $S \sim L$. For $d=2$ lattices, each additional PC generates an additive entropy proportional to $\log{L}$ and thus the additional entropy of  $m$ PC's is $m\log{L}$.  When the two subsystems are connected by $L$ PC's, one recovers the anomalous area law (the 1st term of equation (\ref{area_law})), a connection first pointed out in references \cite{levine_miller}. We argue that the 2nd term of equation (\ref{area_law}), the subdominant $O(L)$ term, may be associated with the single PC (or $0$-dimensional) connection between the two subsystems.

\begin{figure}[ht]
\includegraphics[width=5cm]{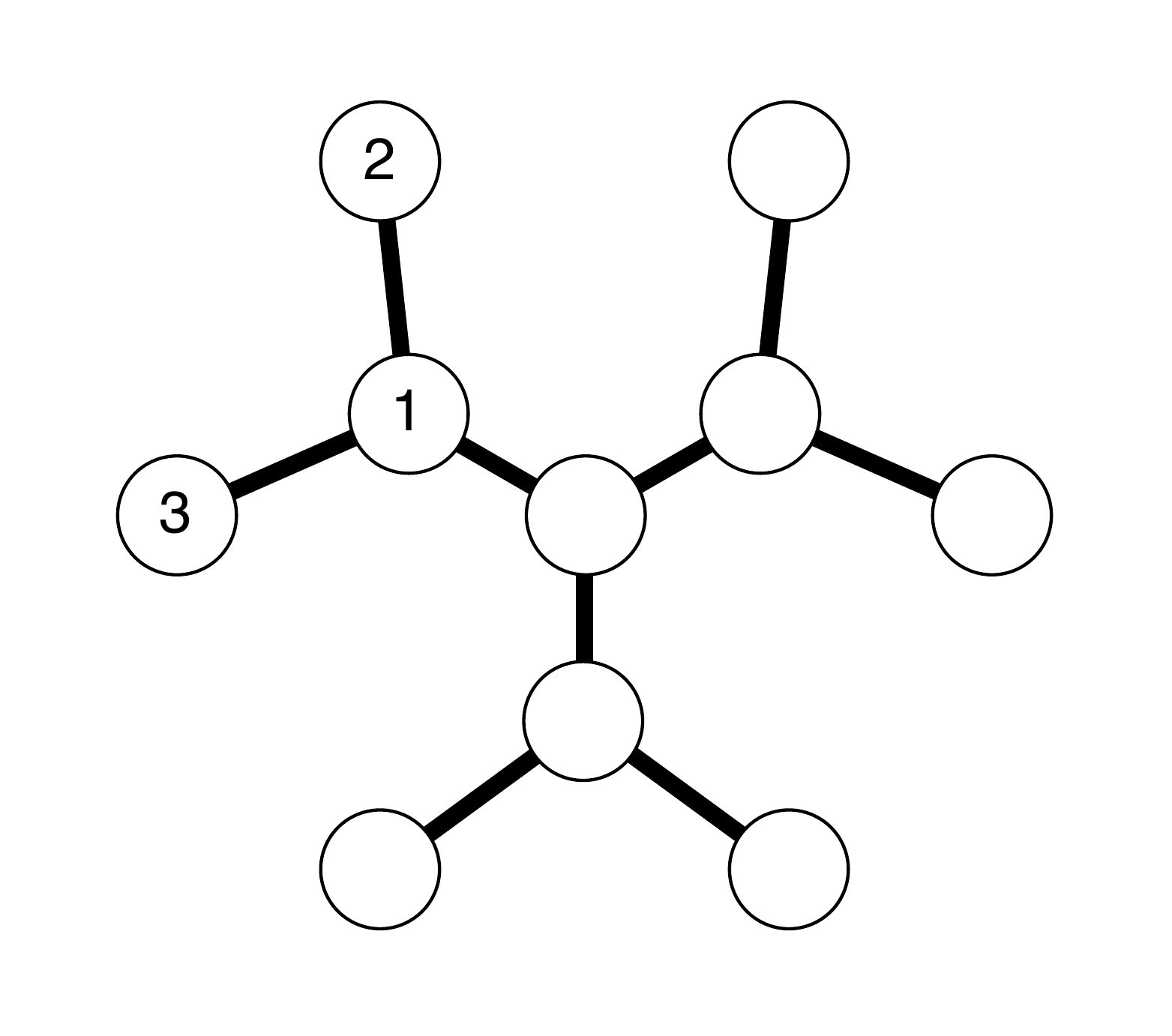}
\caption{A $10$ site $3$ branch site centered Bethe cluster.}
\label{b2}
\end{figure}

Entanglement entropy is decisively influenced by degenerate zero energy states in the single particle spectrum of the combined $A$ and $B$ systems. The origin of the $O(L)$ entropy described here for single PC systems may be traced to the generic $O(L)$ degeneracy of the Fermi surfaces of the disconnected $A$ and $B$ subsystems. We present arguments for two types of lattices (Bethe and translation invariant) demonstrating that most of this exact degeneracy persists when the subsystems are coupled by a single PC.  


Previously it was reported \cite{levine_miller} that in $d=2$ the non degenerate single PC case may be treated by bosonization and mapped to the problem of entanglement in a $1$-dimensional subsystem where the entropy is known. The treatment of the point contacts similar to the x-ray edge or Kondo problem, where the impurity interacts with the $s$-wave sector of the bulk fermion system \cite{GNT}. The PC leads to the entanglement of effectively two (length $L$) 1-d systems and the entropy is thus proportional to $\ln{L}$. The present work clarifies this result, identifying the $O(L)$ entropy associated with a single PC with subleading (bosonic-like) terms in the entropy of the bulk area law.




\section{The model}

Consider two noninteracting systems of spinless fermions (subsystems $A$ and $B$), connected by point contacts spanning the boundary points belonging to set $C$ (fig. 1). The model hamiltonian is:
\begin{eqnarray}
\label{ham}
H &=& H_A + H_B + H_{AB} \\
 &=& -t\sum_{\< i,j \> \alpha=A,B}{(c^{\alpha \dagger}_{i}c^\alpha_{j} + c^{\alpha \dagger}_{j}c^\alpha_{i}})\\
\nonumber&-& y\sum_{k \in C} {(c^{A \dagger}_{k}c^B_{k} + c^{B \dagger}_{k}c^A_{k})}
\end{eqnarray}
where $i$ and $j$ are nearest neighbor site indices and $A$ and $B$ denote the two systems with nearest-neighbor hopping amplitudes $t$ and point contact amplitude, $y$. $c_{i}^{\alpha}$ ($c^{\alpha \dagger}_i$) destroys (creates) fermions at site $i$ in subsytem $\alpha$ and obeys the conventional fermion algebra. The second sum, $H_{AB}$, represents the set of PC's spanning the boundary points belonging to set $C$.  Typically we will consider the case $y=t=1$. Defining the filling fraction $\nu \equiv N_{\rm f}/N_{\rm sites}$, where $N_{\rm f}$ is the number of spinless fermions and $N_{\rm sites}$ is the total number of sites, we will restrict our considerations to the case of $\nu \approx 1/2$. 

We compute the entanglement entropy of subsystem $A$ following the method introduced by Peschel in ref. \cite{peschel_corr_fn}. The entropy may be computed from the eigenvalues of the ground state free fermion correlation matrix
\begin{equation}
\label{corr}
C_{x,y} \equiv \<c_y c^\dagger_x \>
\end{equation}
where $x$ and $y$ are lattice points exclusively within subsystem $A$. Denoting by $\xi_k$ the eigenvalues of $C_{x,y}$, the expression for the entanglement entropy is:
\begin{equation}
\Sent = -\sum_{k}{\left((1-\xi_k)\ln{(1-\xi_k)}+\xi_k \ln{\xi_k}\right)}
\end{equation}

\begin{figure}[ht]
\includegraphics[height=7cm]{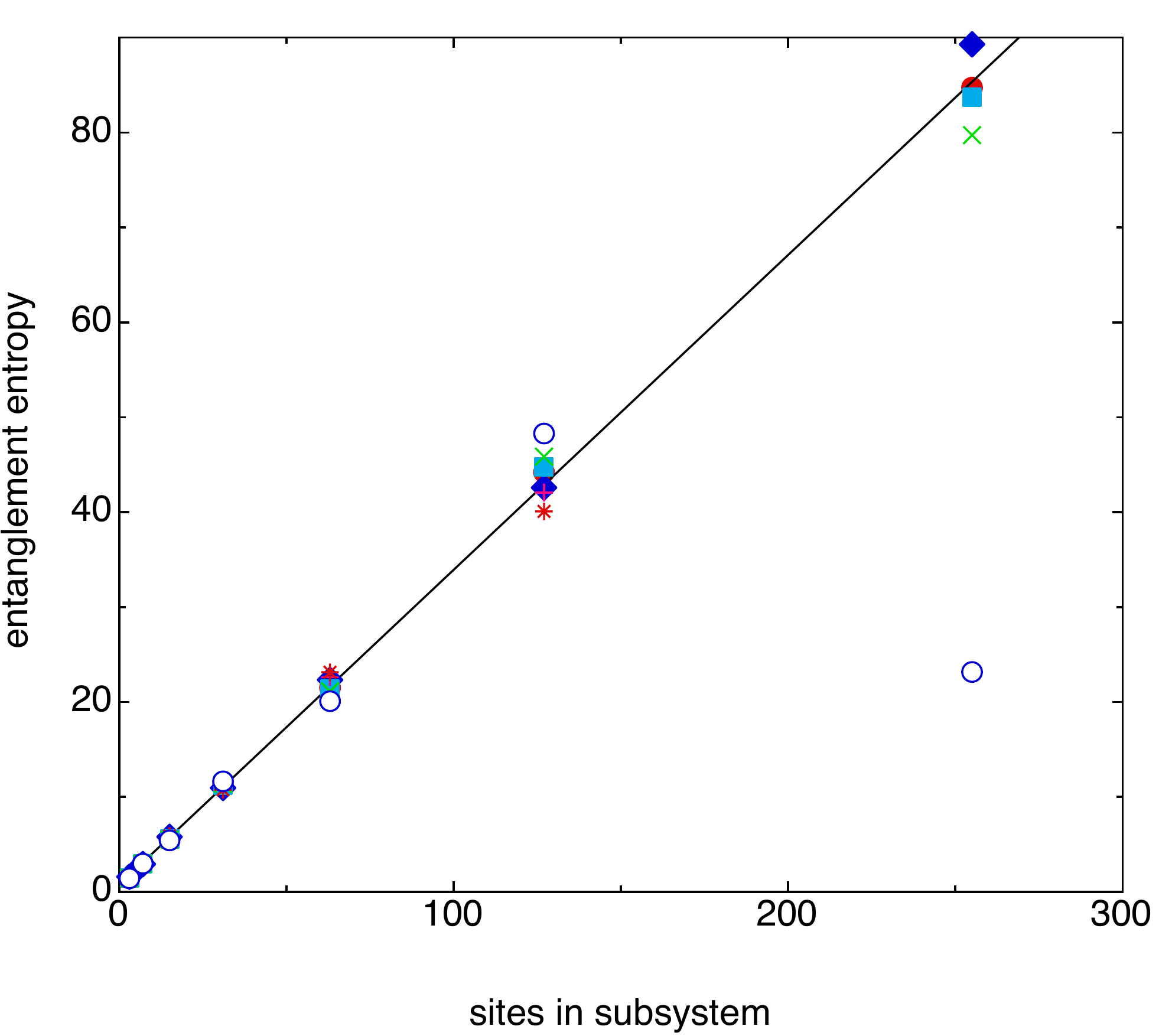}
\caption{Entanglement entropy as a function of interior subsystem size for varying total system sizes, $N$ (bond-centered cluster.)  open circles: $N=510$,  stars: $N=1022$, crosses: $N=2046$, pluses: $N=4094$, squares: $N=8190$, diamonds: $N=16382$, solid circles: $N=32766$.  The line is a linear fit to the results for the $N=32766$ site cluster. }
\label{Sbethe1}
\end{figure}

\begin{figure}[ht]
\includegraphics[height=8cm]{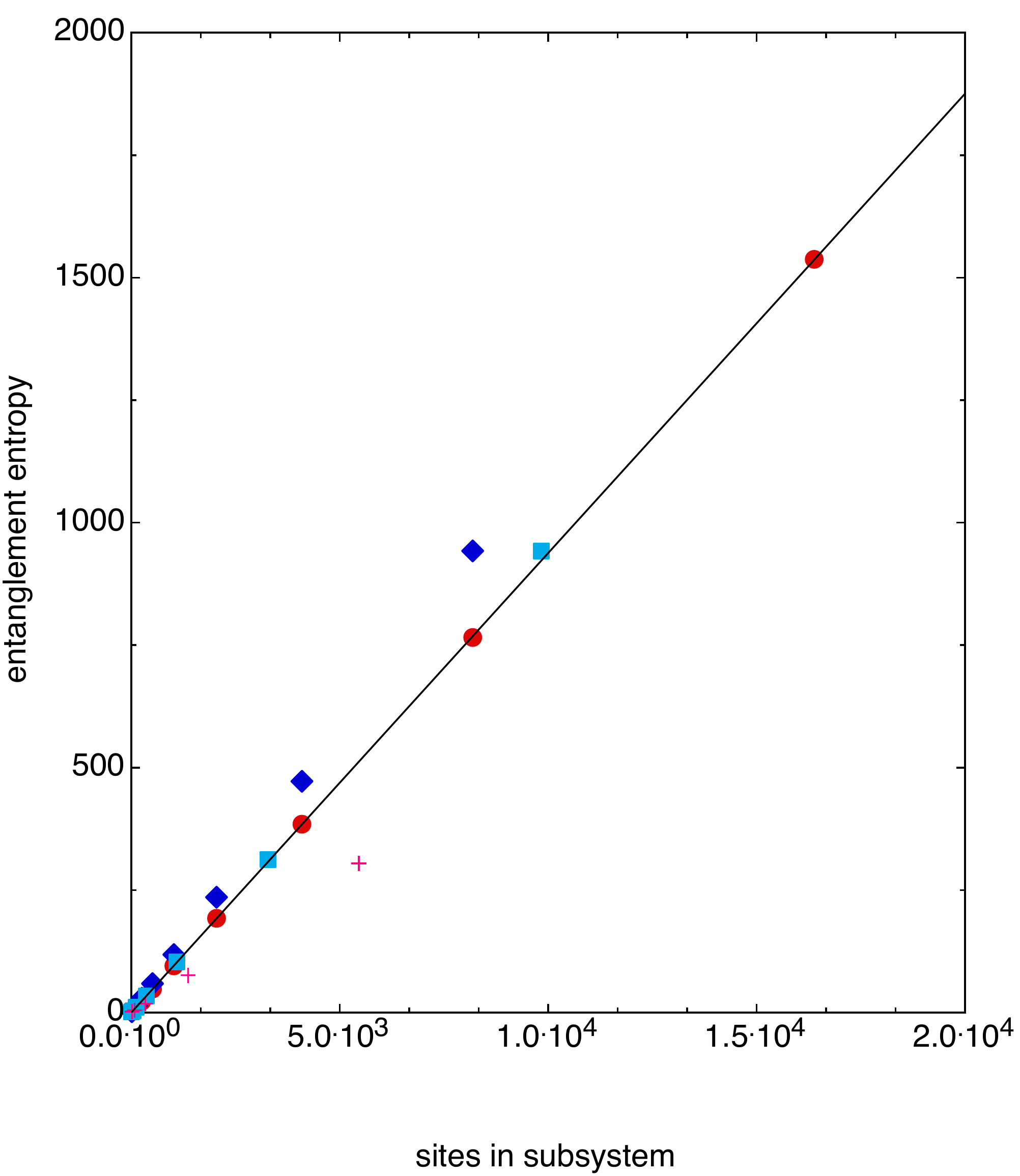}
\caption{Entanglement entropy for half cluster subsystems.  solid circles: 3 leg bond centered, diamonds: 3 leg site centered, squares: 4 leg bond centered, pluses: 5 leg bond centered.  The line is a linear fit to the 3 leg bond centered case.}
\label{Sbethe2}
\end{figure}

\section{Entropy in the Bethe lattice system}

The first example of this behavior we have studied is the entanglement entropy of free fermions on Bethe clusters. Two different types of Bethe clusters have been studied (using the terminology of \cite{bethe}):  bond clustered and site centered clusters. As an illustration, Figure \ref{b1} is a 14 site $z=3$ (3 branch) bond centered cluster while Figure \ref{b2} is a 10 site, site centered cluster. Denoting adjacent sites by an odd/even sublattice designation, note that odd and even sublattices for bond (site) centered clusters are balanced (unbalanced) in their respective number of sites. Thus the bond centered Bethe lattice may be regarded as a particular case of two equivalent ($d<2$) dimensional lattices connected by a point contact.  

\begin{figure}[ht]
\includegraphics[width=7cm]{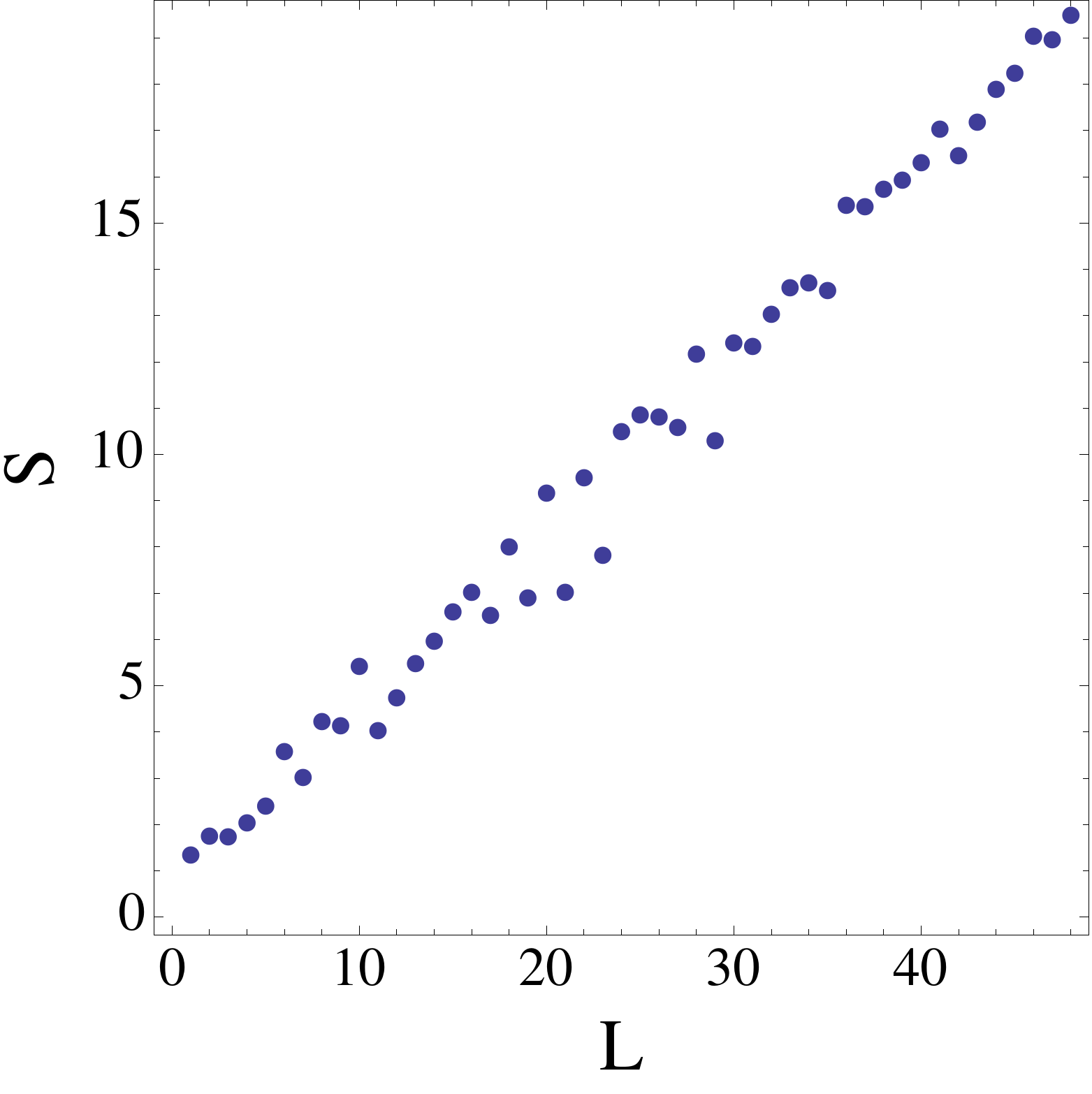}
\caption{\label{fig2} Entropy for a 2-d $L \times L$ lattice (open boundary conditions) with $m=1$ weak links as a function of $L$.}
\label{A}
\end{figure}

Apart from the issue of entropy in weakly coupled systems, the Bethe cluster is of interest from the standpoint of the density matrix renormalization group (DMRG) and tensor network states \cite{Nagaj,Li,Liu}. That is, due to the geometry of the Bethe cluster, it is normal to apply the density matrix renormalization group to the Heisenberg or Hubbard model defined on the Bethe cluster \cite{Otsuka,Friedman,Lepetit,Kumar}. In a typical DMRG approach, the system and subsystem are connected by a single bond, (as in one dimension) hence one would expect that the entanglement entropy would be a constant independent of the size of the system or subsystem. This suggests that DMRG should work extremely well even for large clusters. That is, the number of states to represent a block accurately is roughly $e^{S_{ent}}$ which is a constant independent of the block size. In this respect, the Bethe cluster, at least superficially, resembles one-dimensional systems more closely than two-dimensional systems.

\begin{figure}[ht]
\includegraphics[width=7cm]{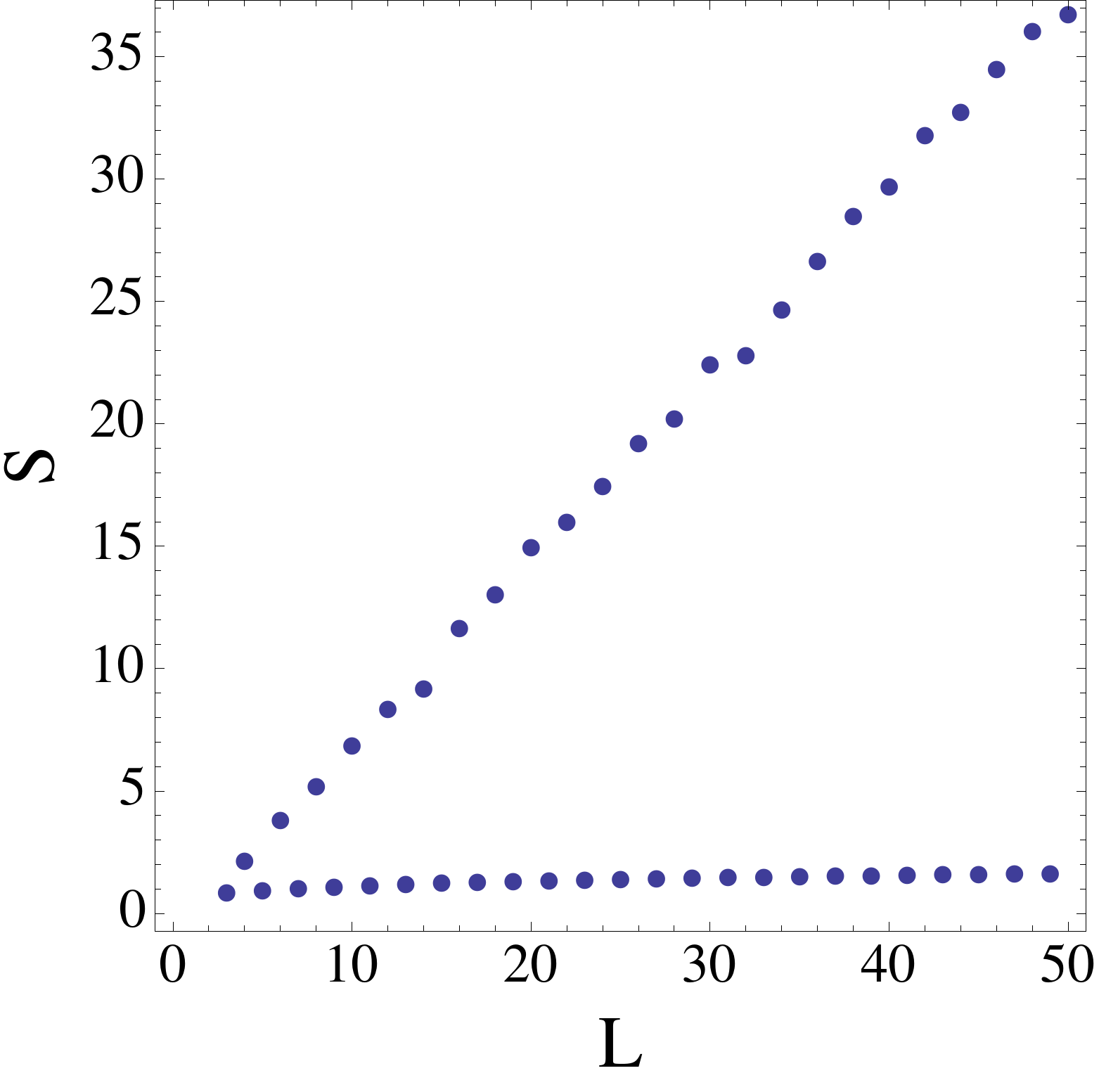}
\caption{Entropy for a 2-d $L \times L$ lattice (periodic boundary conditions) with $m=1$ weak links as a function of $L$.}
\label{B}
\end{figure}

Let us first consider a somewhat different situation from that typically encountered in DMRG namely we take the subsystem inside the system. As an illustrative example, consider in Figure \ref{b1}, a 3 site subsystem, consisting of the sites labelled 1, 2, 3, inside the 14 site bond centered cluster. Using an obvious terminology, the 3 site subsystem shown consists of 2 generations, generation 1 consisting of site 1 and generation 2 consisting of sites 2 and 3. Although clearly this is atypical, it is indicative of the well known behavior of Bethe clusters, that the boundary of the cluster is a finite fraction of the number of sites of the cluster even for large sizes. For the 3 branch cluster and the subsystem of the type shown, a subsystem of $n$ generations has $2^n - 1$ sites and $2^{n-1}$ sites on the boundary, for large subsystems, 1/2 the sites are located on the boundary.

In Figure \ref{Sbethe1}, the entanglement entropy for subsystems of size, $3$ to $255$ (number of generations 2 to 8) is plotted versus the number of sites in the subsystem, for a bond-centered lattice. We are considering the $1/2$ filled case for non interacting spinless fermions. The different symbols and colors correspond to differing system sites ($510$ to $32,766$ sites) and the line is a linear fit to the results for the $32,766$ site system. 


We see from Figure \ref{Sbethe1} that entanglement entropy is roughly a linear function of the numbers of sites in the subsystem and the linearity improves with large systems. This is consistent with the area law (entanglement entropy) of a volume proportional to the boundary area, since for the subsystems considered the number of sites on the boundary is proportional to the number of sites in the subsystem. Note for the 255 site subsystem in the 510 site cluster, the point which lies far off the curve, the boundary points of the subsystem are boundary points of the system.

\begin{figure}[ht]
\includegraphics[width=7cm]{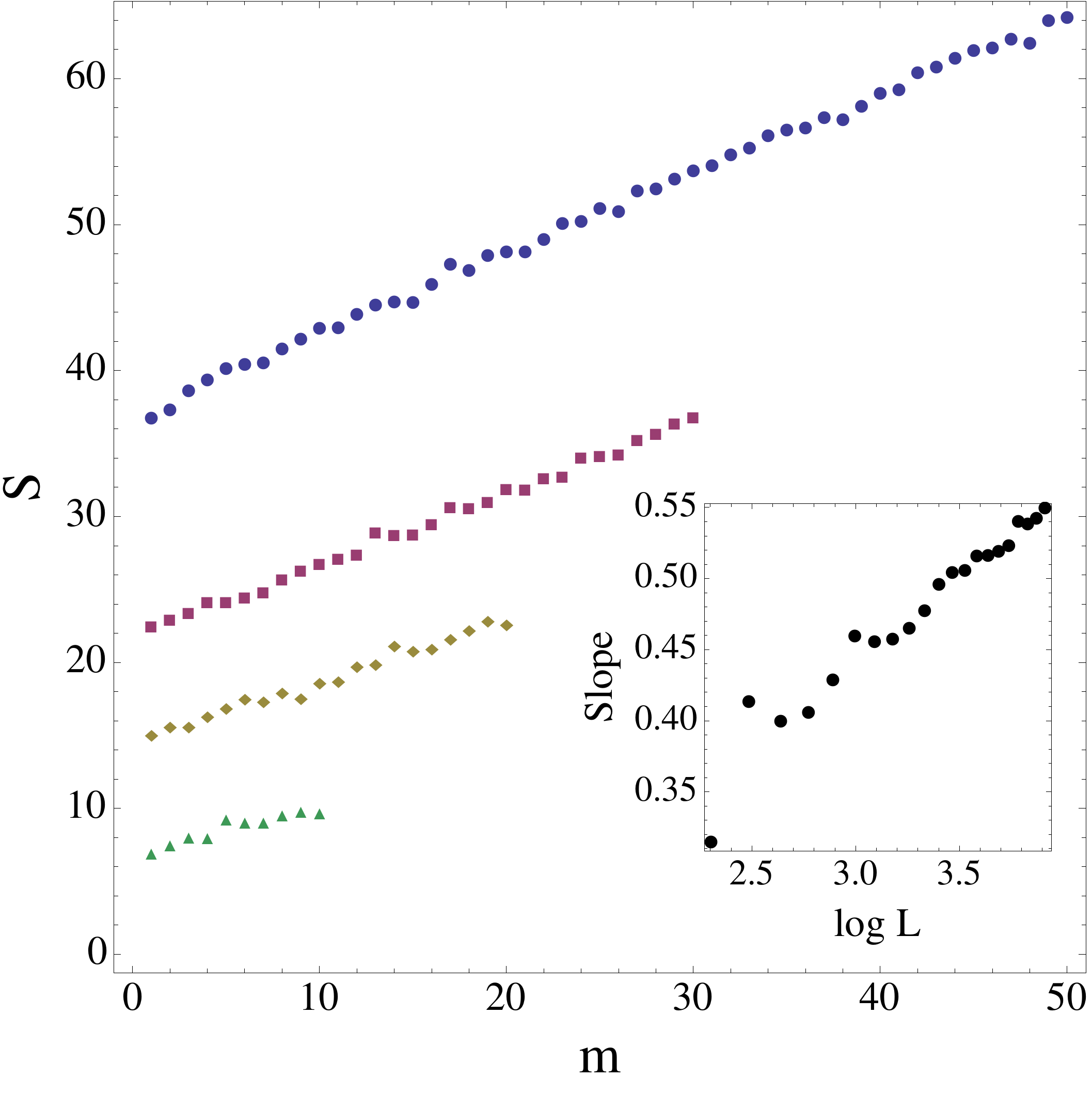}
\caption{Entropies of 2-d $L \times L$ lattices as a function of $m$ weak links for $L=10$ ($ \vartriangle $), $20$ ($ \lozenge $), $30$ ($ \square $), $50$ ($ \circ $). Inset: slopes of these $L$ versus $m$ as a function of $\log L$. Periodic boundary conditions.}
\label{H}
\end{figure}

Let us now consider a situation more similar to that encountered from our previous calculations where a link joins two square square clustered, and to the blocking procedure in DMRG. Take a bond centered cluster (for example Figure \ref{b1}) and pick the subsystem to be the left half of the cluster. This is done in Figure \ref{Sbethe2}, where the entanglement entropy is plotted versus the number of sites in the subsystem. The red dots are for a sequence of 3 legged bond centered clusters ranging in size from 30 to 32,766 in sites. The line is a linear fit to numerical data. We thus see, even though only a single link joins the subsystem to the rest of the system, the entanglement entropy still scales as the number of sites in the subsystem.

The $O(L)$ entropy for the Bethe cluster is traced to a proliferation of zero energy states. It is easy to see for two $n$ generation trees joined by a link (e.g. figure \ref{b1} is two, three-generation trees), there is at least one zero energy state for every two adjacent boundary sites in the cluster. Consider again Figure \ref{b1}. A wave function with an amplitude of $1/\sqrt{2}$ on site 4, $-1/\sqrt{2}$ on site 5 and zero elsewhere is a zero energy eigenstate of the single particle Hamiltonian. Clearly there are 4 such eigenstates for the 14 site cluster. In general, every 2 adjacent boundary sites give rise to one zero energy state. Since for a large cluster of $P$ sites there are of order $P/2$ boundary points, there are at least $P/4$ zero energy states. Hence, unlike the case for the square clusters considered in the next section, one cannot avoid the zero energy states without changing the filling fraction, $\nu$, from 1/2.

\begin{figure}[ht]
\includegraphics[width=7cm]{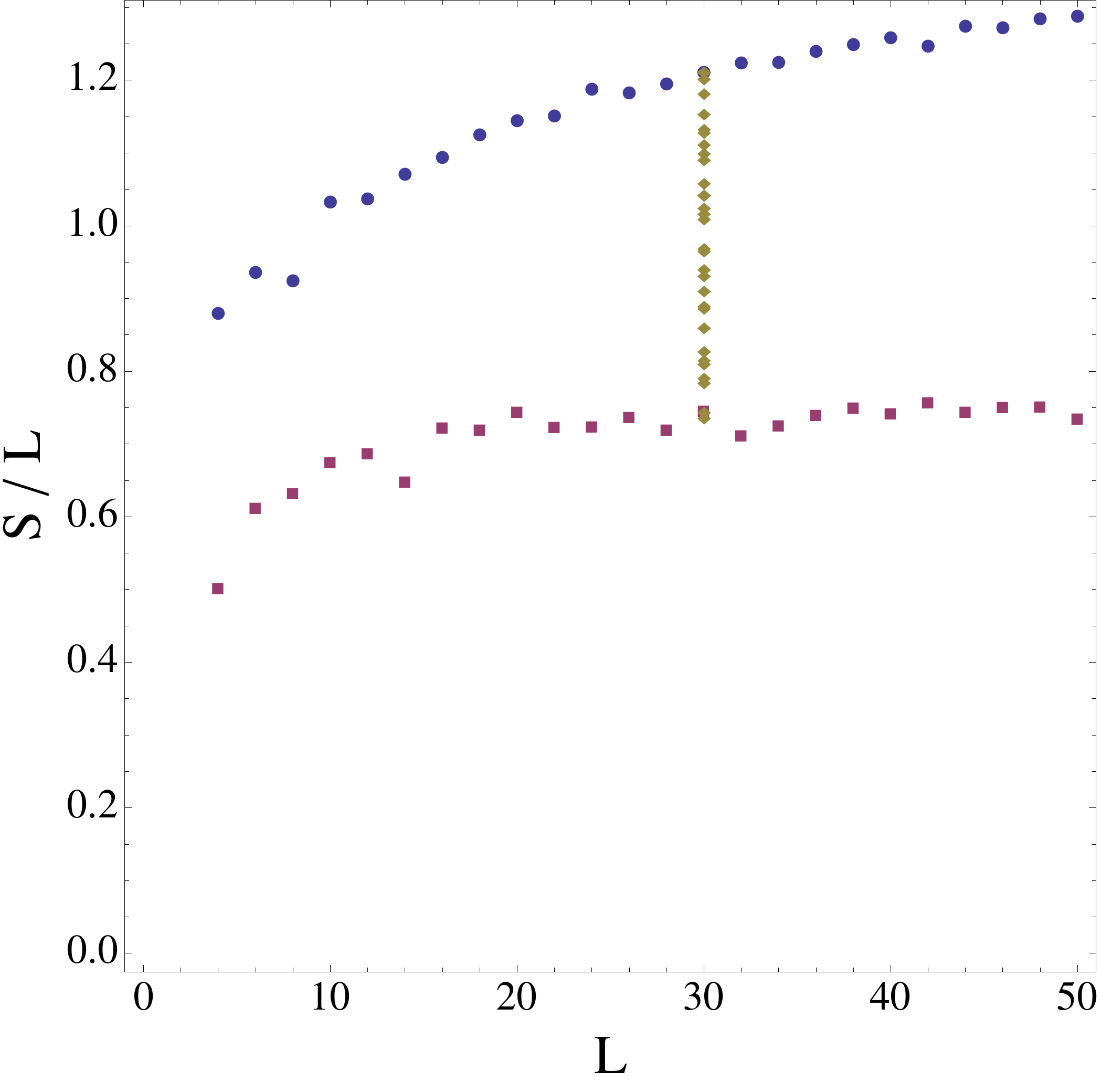}
\caption{Entropies of 2-d $L \times L$ lattices as a function of $L$ connected by $m = L$ ($\circ$) weak links and $m=1$ ($\square$) weak links, periodic boundary conditions. For $L=30$, the explicit dependence of the entropy upon the number of PCs, $m$, is illustrated ($ \lozenge $).}
\label{E}
\end{figure}

In Figure \ref{Sbethe2}, we have also plotted entanglement entropy versus number of sites in the subsystem for the 3 leg site centered Bethe cluster (diamonds) the 4 leg bond centered Bethe cluster (squares) and the 5 leg bond centered Bethe cluster (plus symbols). For the site centered case we took the subsystem to be 1/3 of the cluster (sites 1, 2, 3 in Figure \ref{b2}) while for the bond centered cluster we took 1/2 the cluster as the subsystem. We work at $\nu = 1/2$ and the calculations become more difficult due to the large matrices, with increasing number of legs. In all cases, we find that the entanglement entropy increases linearly with the number of sites in the subsystem, although only a single bond connects the subsystem to the rest of the cluster.

\section{$d=2$ square lattices}

We now study a system of two $d=2$ dimensional lattices connected one or more PC's.  For the numerical part of this study, each subsystem is taken to be a square lattice of linear size $L$ with either periodic or open boundary conditions as shown in Figure \ref{lattice}. As depicted, the subsystems are connected by $m$ point contacts.  For most of the calculations that follow, the filling fraction has been set to $\nu = 1/2$.

For a square lattice of linear size $L$ with open BC's, connected by one PC, Figure (\ref{A}) shows a linear relation between entropy, $S$, and $L$ up to systems sizes of $L=50$. This behavior is similar to what is seen in the Bethe lattice (Figure \ref{Sbethe1}, \ref{Sbethe2}). The comparable calculation for periodic BC's is shown in Figure \ref{B}.  As noted previously \cite{levine_miller}, for odd $L$ the entropies scale logarithmically in $L$; for even $L$ the entropies scale linearly, similar to Figure \ref{A}.

\begin{figure}[ht]
\includegraphics[width=7cm]{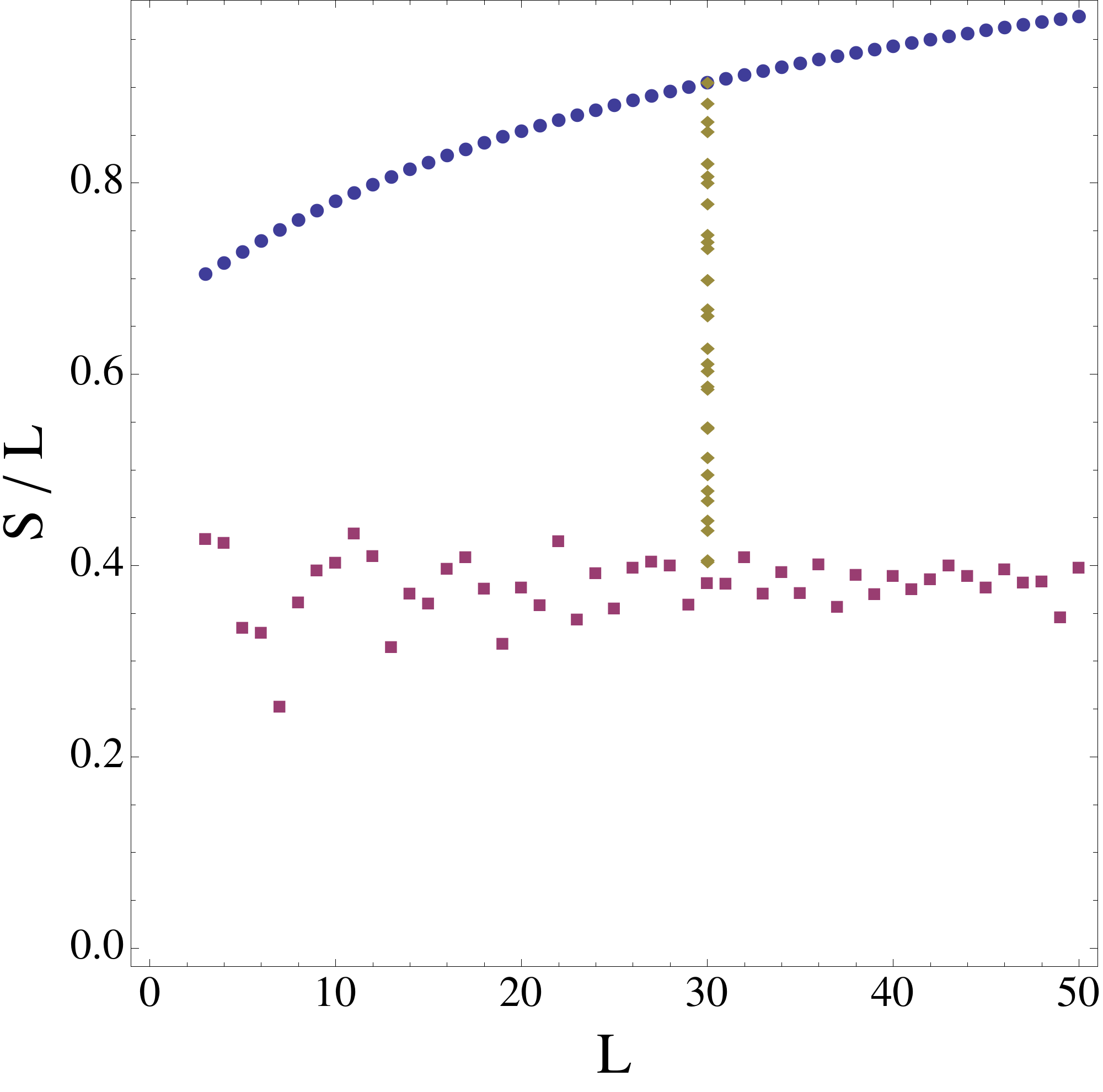}
\caption{Entropies of 2-d $L \times L$ lattices as a function of $L$ connected by $m = L$ ($\circ$) weak links and $m=1$ ($\square$) weak links, open boundary conditions. For $L=30$, the explicit dependence of the entropy upon the number of PCs, $m$, is illustrated ($ \lozenge $).}
\label{F}
\end{figure}

Next we observe the dependence of entropy on the number of PC's connecting the two subsystems. Figure \ref{H} shows the entropy for several lattice sizes, $L$, with a number of PC's ranging from $m=1,L$. Each PC appears to contribute additively to the entropy and the entropy is linear in $m$. Looking at the slope of several such data sets (Figure \ref{H}, inset) each PC appears to contribute an entropy of order $\log{L}$. Thus, this behavior is consistent with anomalous area law, a feature that was first pointed out in references \cite{levine_miller}. Combining the $O(L)$ entropy associated with a single PC, we arrive at an entropy-area law for $d=2$ PC systems:
\begin{equation}
\label{PCarea_law}
S = aL + bm\log{L}
\end{equation}
where $a$ and $b$ are constants and $m$ is the number of PC's.

To clearly differentiate between the $O(L)$ and $O(L\log{L})$ behaviors, the entropies (scaled by $L$) for single and $m=L$ PC systems are plotted as a function of $L$ in figures (\ref{E}, \ref{F}).  Especially for open BC's (Figure \ref{F}) the $m=L$ (complete boundary case) clearly increases monotonically, while the $m=1$ case exhibits roughly constant behavior. Since linearity in $m$ is exhibited in Figure (\ref{H}), the $O(\log{L})$ gap between the curves in figures (\ref{E}, \ref{F}) can only be explained by the PC entropy law (\ref{PCarea_law}). In both figures (\ref{E}, \ref{F}), the proportionality of the entropy to the number of PCs, $m$, is illustrated for the case of $L=30$.

\section{entanglement entropy and single particle degeneracy}

\begin{figure}
\includegraphics[width=7cm]{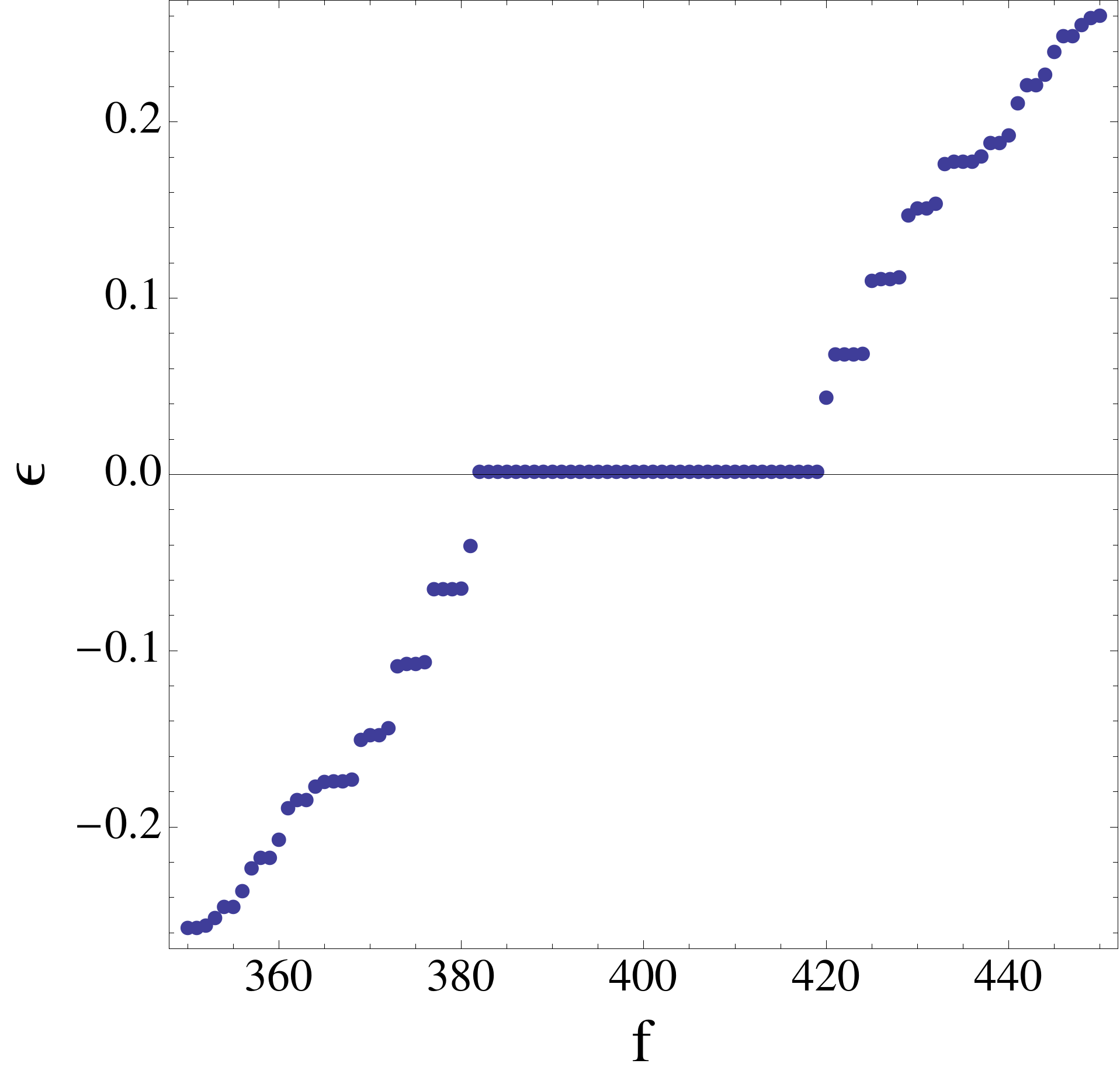}
\caption{Eigenvalues of hamiltonian (Equation \ref{ham}, $y=1$) for $L=20$ 2-d lattice, open boundary conditions. Original $2N$-fold degeneracy $(N=20)$ of the uncoupled $A$ and $B$ subsystems is partially lifted. Two states are separated by a gap of $yN/L^2$ leaving a $2(N-1)$-fold degeneracy, the source of the $O(L)$ entropy. }
\label{D}
\end{figure}

As in the Bethe lattice, the $O(L)$ entropies in the single PC $d=2$ calculations are associated with an $O(L)$ degeneracy in the single particle energy spectrum. The eigenvalues of the hamiltonian (\ref{ham}) are shown in Figure \ref{D} for a typical odd/even $L$. At $\nu=1/2$, the fermi surface of a single disconnected subsystem exhibits an $O(L)$ degeneracy; as seen in Figure \ref{D} this degeneracy appears to (mostly) persist once $A$ and $B$ subsystems are coupled.

For a subsystem with full rotational invariance, the persistence of the degeneracy with one PC is not surprising. Similar to the Kondo problem, the free fermion eigenstates may be written in terms of angular momentum eigenstates.  The PC couples systems $A$ and $B$ through the s-wave state, leaving a sub-extensive, $O(L^{d-1})$, azimuthal angular momentum degeneracy \cite{GNT,levine_miller}.  In our finite system on a lattice, this argument at most applies only approximately for large lattices.  However, the bulk degeneracy is observed to be exactly maintained; specifically, the degeneracy of only one state is removed by the PC, exactly as in the case with full rotational invariance. 

For the present lattice symmetry, we simply state one possible argument that applies to periodic boundary conditions on any size lattice with one PC. Defining the parity operator, $P$, which exchanges $A$ and $B$ subsystems, the eigenstates of free fermions may be written as eigenstates of $P$ since $[P,H_{AB}]=0$. Choosing the PC at site $(0,0)$ the PC part of the hamiltonian may be written in momentum space:
\begin{equation}
\label{PC_ham}
H_{AB} = -\frac{y}{L^2} \sum_{p,p^\prime} (s^\dagger_p s_{p^\prime} - a^\dagger_p a_{p^\prime})
\end{equation}
where $s_p,s_p^\dagger$  ($a_p,a_p^\dagger$) destroy and create symmetric (antisymmetric) $d=2$ lattice momentum states, $s_p = \frac{1}{\sqrt{2}}(c_p^A + c_p^B)$ and $a_p = \frac{1}{\sqrt{2}}(c_p^A - c_p^B)$.  Thus all state within the symmetric (antisymmetric) degenerate subspace are coupled together with the same amplitude $-y/L^2$. An $N \times N$ matrix of the form, $M_{ij}= a\delta_{ij} + b(1-\delta_{ij})$ has a nondegenerate ground state with eigenvalue $a+ (N-1)b$ separated by a gap to a set of $N-1$ degenerate eigenstates with eigenvalues $a-b$. If the original uncoupled $A$ and $B$ systems each exhibit an $N$-fold fermi surface degeneracy (that is, in the parity basis, $N$ symmetric parity states and $N$ antisymmetric parity states), the symmetric and antisymmetric parts of the PC hamiltonian (\ref{PC_ham}) each remove one state, resulting in a persistent $2(N-1)$-fold degeneracy. The nondegenerate states are separated by a gap of $y N/L^2$ as shown in Figure (\ref{D}).

\begin{figure}[ht]
\includegraphics[width=7.0cm]{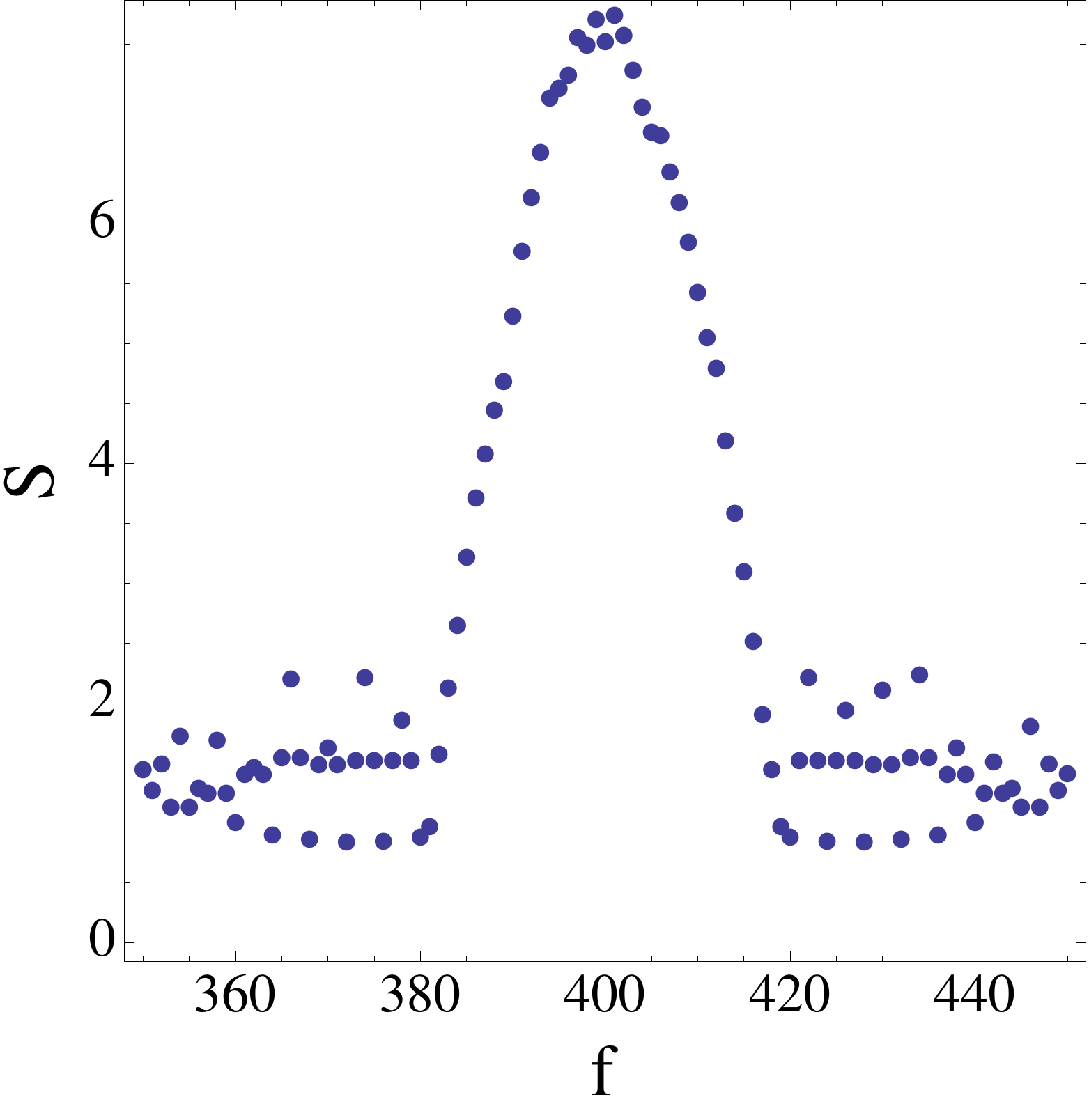}
\caption{\label{fig2} Entropy for 2-d $L \times L$ lattice where $L=20$ with $m=1$ weak links as a function of fermion filling number, open boundary conditions.}
\label{C}
\end{figure}

The dependence of entropy on $\nu$ for an (even) $L=20$ lattice is shown in Figure \ref{C}.  As expected, the entropies remain $O(L)$ until the set of zero energy degenerate states is filled at $\nu = 1/2 \pm O(1/L)$. Similar behavior is seen for even $L$ lattices with periodic boundary conditions. This $O(L)$ entropy behavior should be contrasted with the behavior for odd $L$ lattices with periodic boundary conditions (see Figure \ref{B}).  The latter behavior is associated with fully filling the $O(L)$ degenerate single particle energy levels, a feature of $\nu=1/2$ for these particular lattices.  Odd $L$ generically have an $O(1/L)$ gap at zero energy and entropy correspond to the filled shell condition is $S \sim \log{L}$.

\section{conclusion and outlook}

In summary, we find  for both the Bethe lattice $(d=1)$ and $d=2$ lattices, that a single point contact (PC) connecting two $L^d$ site systems generates an entropy $S \sim L$. For $d=2$ lattices, each additional PC generates an additive entropy proportional to $\log{L}$ and thus the additional entropy of  $m$ PC's is $m\log{L}$.  When the two subsystems are connected by $L$ PC's, one recovers the anomalous area law (the 1st term of equation (\ref{area_law})). Our result also suggests that the subdominant $O(L)$ term  (the 2nd term of equation (\ref{area_law})) may be associated with the degeneracy condition found with a single PC (or $0$-dimensional) connection between the two subsystems.

Turning to the Bethe lattice, what are the consequences of $S \approx O(L)$ for DMRG studies of the Bethe cluster?  In principle, this would imply that it would be difficult to use DMRG for large systems sizes since the number of states to represent a block accurately is $e^{S_{\rm ent}} \approx O(e^{aL})$.  Note that previous DMRG studies of Bethe clusters consider spin systems.  For a two-dimensional ($L \times L$) Heisenberg model, $S \approx L$ (more precisely, $S_2 \approx L$ where $S_2$ is the second Renyi entropy) \cite{Hastings}, while for free fermions $S \approx L \log{L} $.  Hence it is not implausible that Heisenberg models have a different scaling on a Bethe lattice.  

However, a simple argument shows that the valence bond entropy \cite{Alet,Chhajlany} for a Heisenberg model on a bond centered cluster is $S \approx O(L)$.  Divide the bond centered cluster into two parts, dividing along the central bond (i.e. in figure \ref{b1} let a subsystem consist of sites $1,2 \ldots 7$).  In the valence bond basis \cite{Sandvik}, the ground state can be written as linear combinations of bond tilings, where each valence bond connects a site on the even sublattice with the odd sublattice (recall the Bethe clusters are bipartite).  Thus considering a large 3 legged cluster, for 1/2 the cluster, 1/3 (2/3) of the sites are on the even (odd) sublattice.  Hence at least 1/3 of the valence bonds must connect the 2 halves of the cluster leading to an $O(L)$ entropy. 

In contrast, a preliminary calculation of the entanglement entropy of the Heisenberg model on a Bethe cluster---however, using a mean field theory technique successful in higher dimensions \cite{Song}---gives an entropy $S \approx \log{L}$ \cite{Caravan}, compared to the valence bond argument above resulting in $S \approx L$.  It would therefore be of interest to study the entanglement entropy (or $S_2$) on Bethe clusters with unbiased numerical methods, e.g. quantum Monte Carlo \cite{Hastings}.


This research was supported by the Department of Energy DE-FG02-08ER64623---Hofstra University Center for Condensed Matter and Research Corporation CC6535 (GL).

\end{document}